\documentclass{webofc}
\usepackage[varg]{txfonts}   
\usepackage{xcolor}
\usepackage{hyperref}

\providecommand{\sorthelp}[1]{}

\def\lesssim{\mathrel{\hbox{\rlap{\hbox{\lower4pt\hbox{$\sim$}}}\hbox{$<$}}}}
\def\gtrsim{\mathrel{\hbox{\rlap{\hbox{\lower4pt\hbox{$\sim$}}}\hbox{$>$}}}}

\begin{document}
\title{KISS: a spectrometric imager for millimetre cosmology}
%
%

\author{ \lastname{A. Fasano}\inst{1}\fnsep\thanks{\email{alessandro.fasano@neel.cnrs.fr}}
\and M. Aguiar\inst{2} \and A. Benoit\inst{1} \and A. Bideaud \inst{1} \and O. Bourrion\inst{3} \and M. Calvo\inst{1} \and A. Catalano\inst{3} \and A. P. de Taoro\inst{2} \and G. Garde\inst{1} \and A. Gomez\inst{4} \and M. F. Gomez Renasco\inst{2} \and J. Goupy\inst{1} \and C. Hoarau\inst{3} \and R. Hoyland\inst{2} \and J. F. Mac\'ias-P\'erez\inst{3} \and J. Marpaud\inst{3} \and A. Monfardini\inst{1} \and G. Pisano\inst{5} \and N. Ponthieu\inst{6} \and J. A. Rubi\~no Mart\'in\inst{2} \and D. Tourres\inst{3} \and C. Tucker\inst{5} \and A. Beelen \and G. Bres\inst{1} \and M. De Petris\inst{8} \and P. de Bernardis\inst{8} \and G. Lagache\inst{7} \and M. Marton\inst{2} \and R. Rebolo\inst{2} \and S. Roudier\inst{3}
}

\institute{Institut Néel, CNRS and Université Grenoble Alpes, 25 av. des Martyrs, F-38042
	Grenoble, France
\and
			Instituto de Astrofísica de Canarias, Vía Láctea s/n, E-38205 La Laguna, Tenerife, Spain
\and
           	Univ. Grenoble Alpes, CNRS, Grenoble INP, LPSC-IN2P3, 53, avenue des Martyrs, 38000 Grenoble, France
\and
			Centro de Astrobiologia (CSIC-INTA), Madrid, Spain
\and
			Astronomy Instrumentation Group, University of Cardiff, UK
\and
			Univ. Grenoble Alpes, CNRS, IPAG, F-38000 Grenoble, France 
\and
			Aix Marseille Université, CNRS, LAM (Laboratoire d'Astrophysique de Marseille) UMR 7326, 13388, Marseille, France
\and
			Dipartimento di Fisica, Sapienza Universit\`a di Roma, Piazzale Aldo Moro 5, I-00185 Roma, Italy
          }

\abstract{%
Clusters of galaxies are used to map the large-scale structures in the universe and as probe of universe evolution. They can be observed through the Sunyaev-Zel'dovich (SZ) effect. In this respect the spectro-imaging at low resolution frequency is an important tool, today, for the study of cluster of galaxies. We have developed KISS (KIDs Interferometer Spectrum Survey), a spectrometric imager dedicated to the secondary anisotropies of the Cosmic Microwave Background (CMB). The multi-frequency approach permits to improve the component separation with respect to predecessor experiments. In this paper, firstly, we provide a description of the scientific context and the state of the art of SZ observations. Secondly, we describe the KISS instrument. Finally, we show preliminary results of the ongoing commissioning campaign.
}
\maketitle	
\section{Scientific context}
\label{introduction}

The SZ effect is a spectral distortion of the CMB due to the interaction between CMB photons and the hot electrons in cluster of galaxies (see \cite{Sunyaev_Zeldovich:SZE_forecast_original_paper} for a detailed description). This effect is exploited for several purpose in contemporary cosmology: the detection of clusters of galaxies, their morphological mapping and the evaluation of the cosmological parameters from the $\Lambda CDM$ model \cite{lCDM}.

The overall SZ effect includes two components:
the thermal SZ (tSZ) due to the Inverse Compton scattering with the Intra Cluster Medium (ICM) in clusters and the kinematic SZ (kSZ) induced by its celestial structure bulk motion.

The tSZ effect is particularly interesting as its spectral signature provides a unique possibility to detect and study clusters of galaxies at any redshift, because it does not have cosmological dimming. Furthermore, the tSZ has been proved to be a good proxy of
the mass of the cluster showing lower dispersion with respect to the precursor optical, X-rays or lensing observations. However, component separation and astrophysical contamination by foreground emission (e.g. thermal dust, synchrotron, dusty and radio galaxies) limits the accuracy of the reconstruction of the tSZ effect (see \cite{ruppin} for more details): we require, thus, a multi-wavelength approach.

On a chronological point of view, the first SZ surveys have been acquired with the South Pole Telescope SPT \cite{Staniszewski_2009:first_cluster_discovery_with_SZE}, the Atacama Cosmology Telescope ACT \cite{Kosowsky_2003:the_atacama_cosmology_telescope_ACT} and the Planck \cite{Hurier_2014_CMB_temperature_evolution_from_Planck} satellite. They are affected by relatively poor angular resolutions (from 5' for Planck to 1' for SPT) and are, thus, not able to resolve cluster sub-structures. The efforts on many incoming SZ experiments are oriented on reaching angular resolution down to 20'' to realise detailed clusters maps. This goal is reached by exploiting high resolution millimetre telescopes with new generation instrumentations, like with NIKA \cite{Catalano2014}, NIKA2 \cite{nika2} and MUSTANG-2 \cite{mustang2}, reached by exploiting high resolution millimetre telescopes with new generation instrumentations.

However, so far, the main limitation has been the number of observing frequencies. Indeed, with sufficiently precise spectroscopy measurements we could independently measure the tSZ (including relativistic corrections) and the kSZ, which would give access with higher accuracy to the mass (tSZ flux), the cluster proper motion along the line-of-sight and temperature (relativistic corrections to the tSZ). This is the purpose of KISS experiment and OLIMPO (see \cite{olimpo} for an overview). The spectral resolution demanded for such a scientific target is of few GHz with a total bandwidth from 10 up to 100 GHz (see \cite{pdb} for details), but an high mapping speed is required to cover the typical low-redshift cluster of galaxies size (few degrees). This is the main reason for adopting a wide Field of View (FoV) Martin Puplett Interferometer (MPI) (see \cite{MPI} for a detailed description of the MPI and \cite{ltd18} for the KISS configuration) with large-array detectors with respect to the heterodyne at high spectral resolution (such as ALMA \cite{ALMA} and NOEMA \cite{NOEMA}) covering a smaller instantaneous FoV.

\section{KISS}
\label{kiss}

The main scientific objective of KISS is to perform a follow-up of low redshift clusters
previously detected by the Planck satellite, in order to improve the estimate of the
tSZ flux, which is directly linked to the mass of the cluster. Since the main objective of the experiment is the overall tSZ flux, we have adopted only moderate improvement on the angular resolution with respect to Planck. 

\begin{figure}[ht]
	\centering
	\includegraphics[scale=0.45]{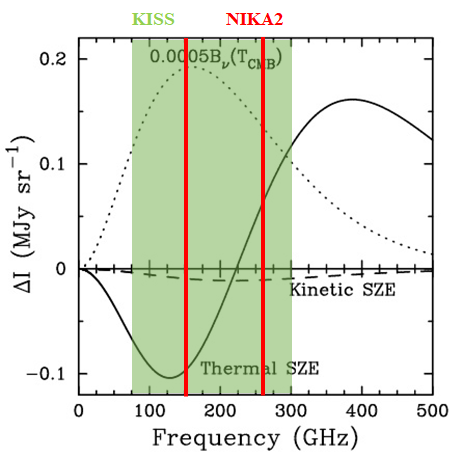}
	\caption{Spectral distortion intensity of the CMB radiation due to SZ effect. The solid line represents the thermal SZ and the dashed line is the kinetic SZE. The black body spectrum of the CMB scaled by a factor 0.0005 is shown by the dotted line. The cluster properties	used to calculate the spectra are an electron temperature of 10 keV, a Compton $y$ parameter of 10$^{-4}$, and a peculiar velocity of 500 km s$^{-1}$ \cite{carl}. It is reported KISS (spectrometer) frequency coverage and the central NIKA2 frequencies (photometer).}
	\label{fig:sze}
\end{figure}

The experiment is ground-based and, thus, affected by atmospheric limitation: the transmission performance at high frequency is degradated. We, thus, observe in the frequency range 80 to 300 GHz. In terms of spectral resolution the complexity of the foreground emission requires measurements at multiple frequencies to separate them from the SZ signal. Although multiple photometric channel experiments can partially fulfil this requirement (this was the case for the Planck HFI instrument with seven useful channels), a spectrometer will give the possibility to adapt the spectral resolution to the specific observation details. Therefore, we have built a MPI with spectral resolution up to 1 GHz. It exploits two arrays of Kinetic Inductance Detectors (KID) (see \cite{kid} for an overview), camera (632 pixels in total) cooled to 150 mK by a custom dilution refrigerator. In figure~\ref{fig:sze} shows the frequency coverage of KISS, and display the central frequencies of the 2 NIKA2
bands for comparison.

\subsection{Laboratory characterisation}

\begin{figure}[ht]
		\centering
		\includegraphics[scale=0.24]{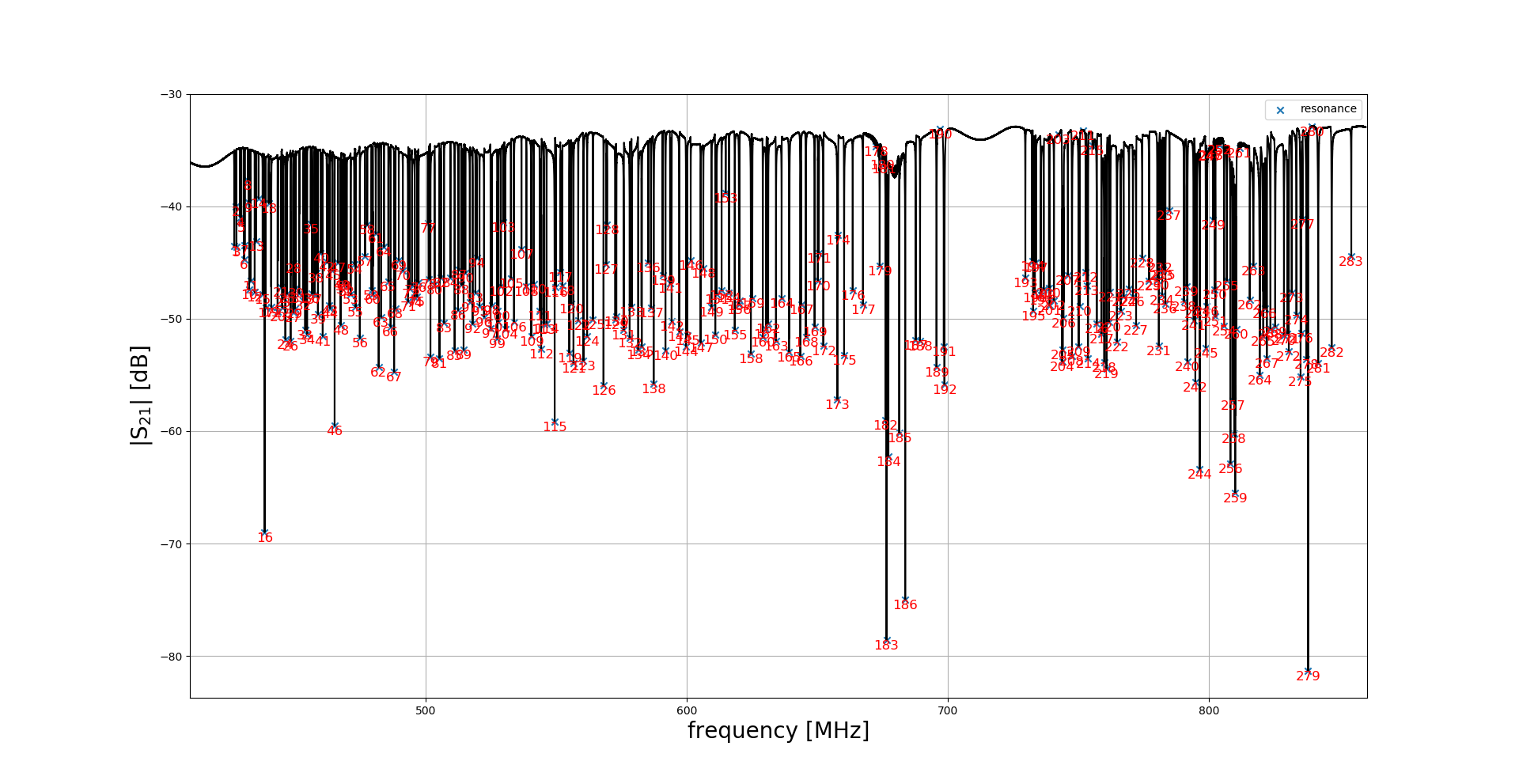}
		\caption{Vector Network Analyser scan for one of the KID array. $|S_{21}|$ scattering parameter modulus vs bias line frequency.}	
		\label{fig:vna}
\end{figure}

The necessary condition to fully validate the instrument and to proceed to the telescope integration has been the array characterisation. In figure~\ref{fig:vna} we see the frequency response of the KIDs for one of the KISS array: it is the representation of the reflecting $|S_{21}|$ scattering parameter. Each pixel (KID) needs an excitation tone injected so, in terms of bias line modulus, an $|S|$ [dBm] signal: $|S_{21}|$ is the output/input modulus ratio. The KID is a resonator, it absorbs energy at its resonant frequency. Each vertical line in the figure corresponds to a pixel. Not all the 316 pixel are perfectly working and the resonance fitting identifies 283 of them. The break at 700 MHz is due to the different capacitor geometry imposed by the design: in a LEKID (see \cite{kid} for the details) the interdigital capacitor fingers are cut to vary the resonance frequency. When the capacitor design reaches the limit, i.e. the interdigital fingers are all cut,  the capacitor design is changed. The  sinusoidal shape is caused by mismatching of the bias line with the 50 $\Omega$ RF lecture line.

\begin{figure}[ht]
		\centering
		\includegraphics[scale=0.15]{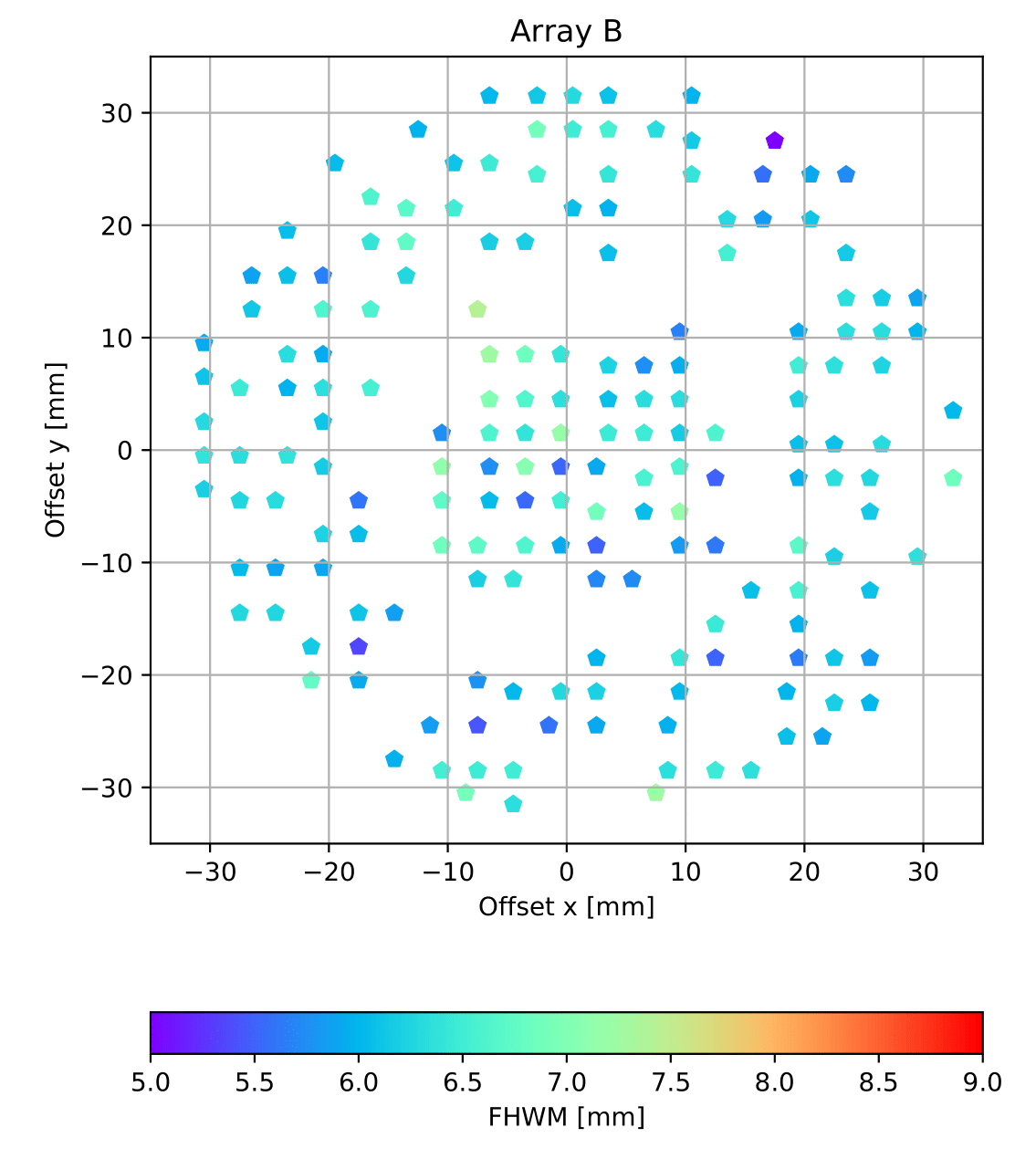}
		\caption{Geometry characterisation of the array: each point is a pixel and the axes are the physical position in the x-y plane. The void regions are due to a defect in the electronics that cannot inject the 316 tones required, it will be a future improvement.}
		\label{fig:geometry}
\end{figure}

\noindent There is a mismatch between the resonance frequencies expected from calculations based on the KID design geometry and the actual frequencies observed in practice. This is probably due to an imperfect control of the fabrication process, the resonance frequency being very sensitive to small variations in the KID meander thickness and width. It is, thus, necessary to adopt an empirical method  to map the pixels on focal plan. For this purpose we simulate a typical sky background through a 30 K black-body (the atmosphere is approximately a 300 K black-body with 0.1 emissivity at mm-wavelengths) and we use a room-temperature pointing source to simulate the passage of a source in the KID array FoV. There is a univocal relation between the sky coordinates of the source and the position of its image on the array, hence the passage of the source image on a KID allows identifying its corresponding tone and position on the array. In figure~\ref{fig:geometry} we see the laboratory result for the KISS focal plane characterisation in terms of position and FWHM of the beam.

\subsection{Integration to the telescope}

\begin{figure}[ht]
		\centering
		\includegraphics[scale=0.45]{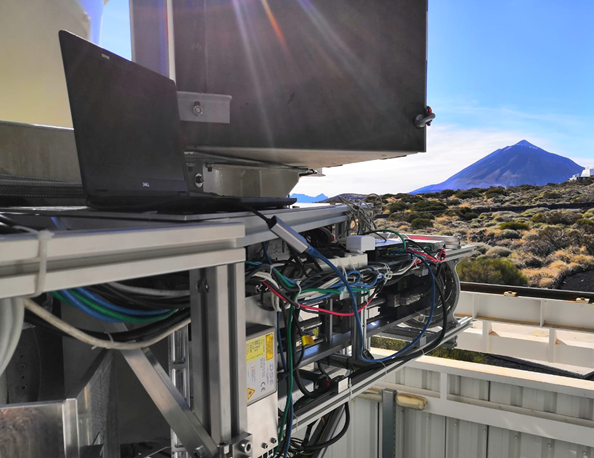}
		\caption{KISS instrument integrated on the QUIJOTE telescope in Tenerife. On the backgroud the Teide volcano.}		
		\label{fig:kiss}
\end{figure}

KISS is installed at the 2.25 m QUIJOTE telescope of the Teide Observatory in Tenerife, since February 2019. In figure \ref{fig:kiss} we see the instrument integrated at the telescope. For more details on the instrument conception and design see \cite{ltd18}.

\section{Preliminary commissioning results}
\label{preliminary}

\begin{figure}[ht]
	\centering
	\includegraphics[scale=0.6]{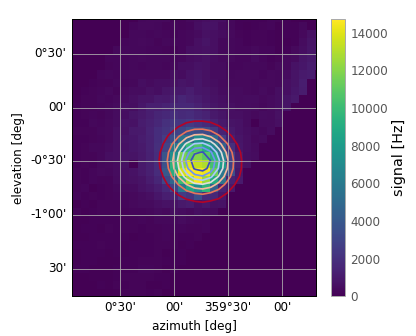}
	\caption{Photometric mapping of the Moon. Azimuthal-Elevation map in hertz. These are single pixel data for a 3x3 degrees map. The contour lines represent the 2-d gaussian fit of the Moon from 1 to 7 sigma.}
	\label{fig:pixmap}
\end{figure}

The first purpose of the commissioning campaign was to validate the instrument on its multi-frequency capability, to asses the feasibility and to proceed to the instrument characterisation on the sky. Before a multi-frequency exploitation of the instrument, we have started with a standard photometric approach, i.e., taking the signal all over the band. In figure~\ref{fig:pixmap} we see the Moon observation.

\begin{figure}[ht]
	\centering
	\includegraphics[scale=0.5]{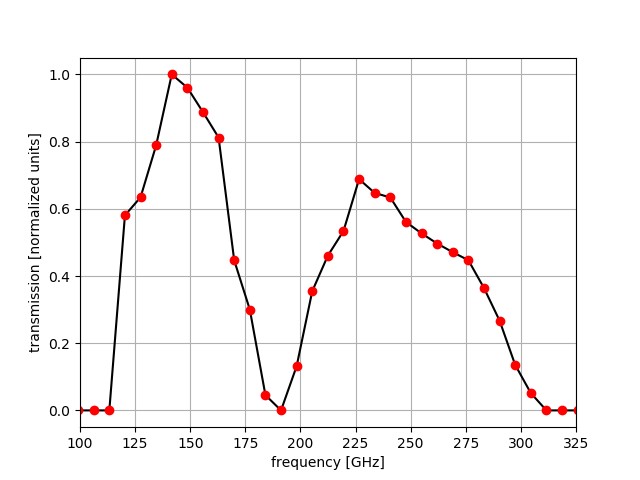}
	\caption{Normalised transmission spectrum of a KISS array. It is performed with a flat-field observation, i.e. a diffuse black body at 300 K. It is RJ corrected and presents a 6 GHz bin.}
	\label{fig:band}
\end{figure}

Starting from the photometric detection, we characterised the transmission spectrum at the telescope. The result is reported in figure~\ref{fig:band}, which is obtained with a black-body at 50 K as the secondary MPI source. Since the calibrator is a 300 K diffuse flat field the transmission spectrum is corrected to take into account the increasing Rayleigh-Jeans (RJ) spectral radiance $\left(\propto\nu^2 \right)$.  We corrected the transmission spectrum by a factor $\nu^2_0/\nu^2$, where $\nu_0$ represents the 150 GHz reference frequency. With such a correction the incoming spectral radiance becomes flat, i.e., it depends just on the temperature.\footnote{$\tilde{H}_\nu \doteq \left[ H_\nu \cdot S^{calib}_{\nu} \right]\cdot\nu_0^2 / \nu^2$, where $\tilde{H}_\nu$ is the RJ corrected transmission spectrum, $H_\nu$ is the transmission spectrum, $S^{calib}_{\nu}$ is the incoming spectral radiance $\left(\propto\nu^2 \right)$ and $\nu_0$ the 150 GHz reference frequency.} We obtained this result making a mean distribution over all the pixels.

\begin{figure}[ht]
		\centering
		\includegraphics[scale=0.12]{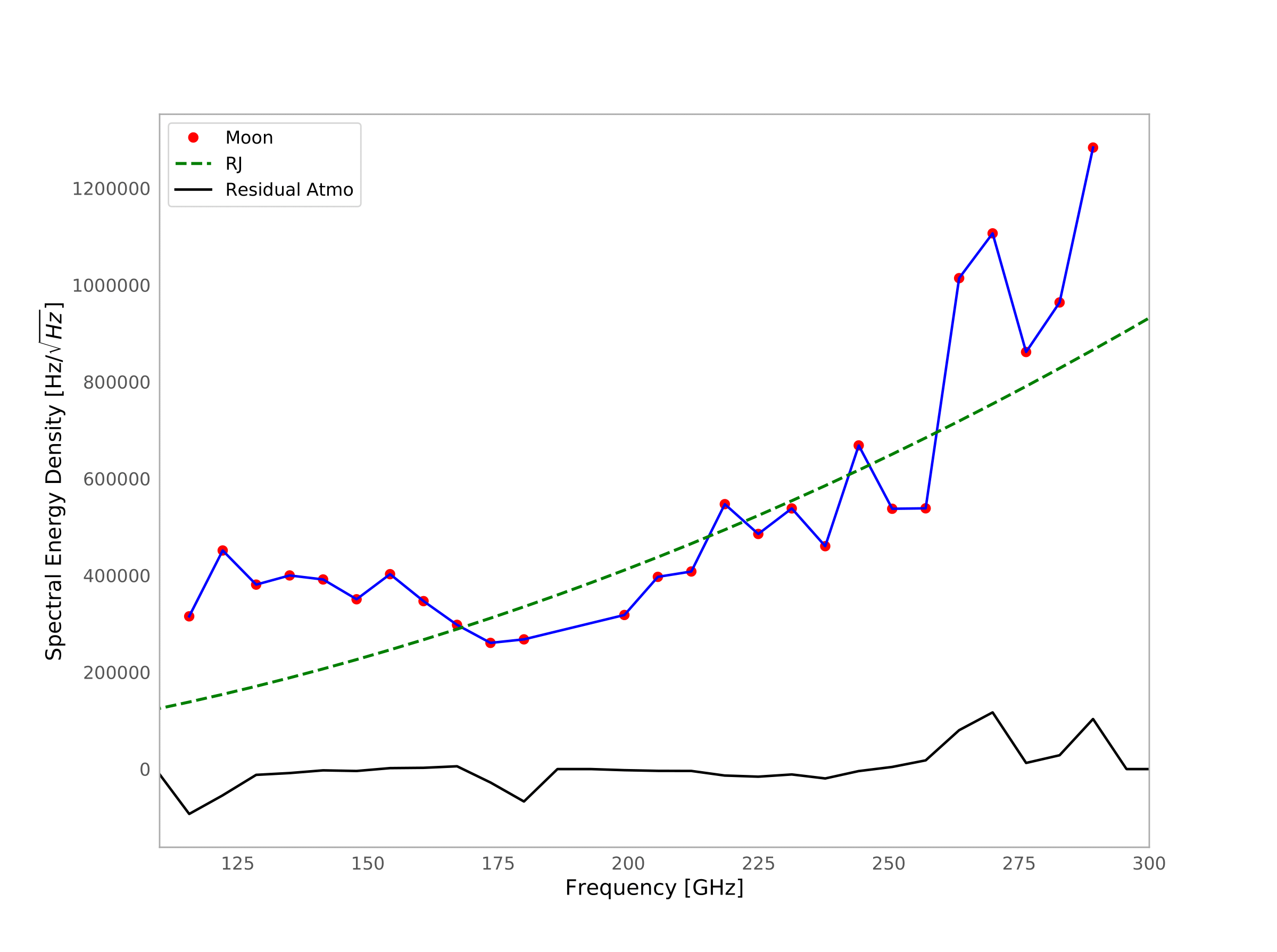}
		\caption{Spectral energy density of the Moon. The plot displays the Moon spectrum in red dots with 6 GHz bin, the RJ curve in green dashed line and the atmospheric residuals in black line. The notch filter cuts the 180 GHz bin. The observation has been performed with a diaphragm to not saturate the detectors.}
		\label{fig:sed}
\end{figure}

\noindent The plotted transmission shows peculiar features: it is down-limited by absorber material (aluminium)\footnote{The minimum detectable frequency ($\nu_{min}$), aka cut-off frequency, is defined by the critical temperature ($T_c$) of the absorber, by the equation: $2h\nu_{min}=2\Delta(T= 0 \text{ kelvin})= 3.5 k_{B} T_c$, where $h$ is the Planck constant and $k_B$ is the Boltzmann constant. See \cite{kid} and \cite{gao} for further details.}, upper limited by the optical filtering and it has the 180 GHz cut, obtained by a notch filter, to avoid the H$_2$O atmospheric absorption line. We have planned, for the future, to replace the aluminium absorber with a bi-layer titanium-aluminium to reach 80 GHz (see \cite{bilay} for the bi-layer technology). In figure \ref{fig:sed} we see the Spectral Energy Density of the Moon convoluted with the transmission spectrum.

\section{Conclusion and perspectives}
In this work we have demonstrated the multi-frequency capability of the KISS instrument; in particular we characterised the focal plane in the laboratory and we have coupled it to an actual millimetre telescope, characterised the transmission spectrum and mapped the Moon with a spectral resolution meeting our expectations.

Although KISS is, standalone, capable to produce scientific results it is, furthermore, a pathfinder for future studies. It would be possible to use its technological concept and install a similar instrument to a telescope with improved angular resolution to aim clusters at higher redshift. This is the main idea of CONCERTO \cite{concerto}: a future experiment that will exploit the expertise developed for KISS. The perspective for KISS is to fully characterise the instrument on sky and observe cluster of galaxies.

%
%
\bibliography{bibliography.bib}

\end{document}